\documentclass[runningheads]{llncs}

\usepackage[T1]{fontenc}
\usepackage{graphicx} \usepackage{amsmath}
\usepackage{bm}
\usepackage{listings}
\usepackage{orcidlink}
\usepackage{hyperref}
\usepackage{multirow}
\usepackage{tabularx}
\usepackage{float}
\usepackage[numbers]{natbib}
\usepackage{color, colortbl}
\usepackage{algorithm,algpseudocode}
\usepackage[colorinlistoftodos]{todonotes}
\usepackage{pifont}
\newcommand{\xmark}{\ding{55}}
\PassOptionsToPackage{hyphens}{url}
\usepackage[hyphenbreaks]{breakurl}
\usepackage{xcolor}
\usepackage{tikz}
\usepackage{svg}
\usetikzlibrary{chains,automata,positioning, shapes, arrows.meta,matrix,calc}

\usepackage{color,xspace,listings,comment,tabulary}

\definecolor{light-grey}{gray}{0.95}
\definecolor{light-gray}{gray}{0.95}
\definecolor{codegreen}{rgb}{0,0.6,0}
\definecolor{codegray}{rgb}{0.5,0.5,0.5}
\definecolor{codepurple}{rgb}{0.58,0,0.82}
\definecolor{backcolour}{rgb}{1,1,1}

\lstdefinestyle{mystyle}{
  backgroundcolor=\color{light-grey},   commentstyle=\color{codegreen},
  keywordstyle=\color{magenta},
  numberstyle=\tiny\color{codegray},
  stringstyle=\color{codepurple},
  basicstyle=\ttfamily\scriptsize,
  breakatwhitespace=false,
  escapeinside={(*}{*)},
  breaklines=true,                 
  captionpos=b,                    
  keepspaces=true,                 
  numbers=left,                    
  numbersep=5pt,                  
  showspaces=false,                
  showstringspaces=false,
  showtabs=false,                  
  tabsize=2
}

\algnewcommand{\algorithmicand}{\textbf{ and }}
\algnewcommand{\algorithmicor}{\textbf{ or }}
\algnewcommand{\OR}{\algorithmicor}
\algnewcommand{\AND}{\algorithmicand}

\definecolor{flbl}{HTML}{a1bbf6}

\usepackage[normalem]{ulem}

\tikzset{
  treenode/.style = {shape=rectangle, rounded corners,
                     draw, anchor=center,
                     text width=5em, align=center,
                     inner sep=1ex},
  root/.style     = {treenode, font=\ttfamily},
  env/.style      = {treenode, font=\ttfamily\normalsize},
  finish/.style   = {root},
}

\begin{document}
\newtheorem{myprop}{Property}

\title{An SMT Formalization of Mixed-Precision Matrix Multiplication}
\subtitle{Modeling Three Generations of Tensor Cores}
\author{Benjamin Valpey\inst{1} \and
Xinyi Li\inst{2} \and
Sreepathi Pai\inst{1} \and Ganesh Gopalakrishnan \inst{2}}
\authorrunning{B. Valpey et al.}
\institute{University of Rochester, Rochester, NY 14627, USA \email{\{bvalpey,sree\}@cs.rochester.edu}\\ \and University of Utah, Salt Lake City, UT 84112, USA \email\{xin\_yi.li@utah.edu, ganesh@cs.utah.edu\}
}

\date{}
\maketitle

\begin{abstract}
Many recent computational accelerators provide non-standard (e.g.,
reduced precision) arithmetic operations to enhance
performance for floating-point matrix multiplication.  Unfortunately, the properties of these accelerators are
not widely understood and lack sufficient descriptions of their behavior. This makes it difficult for tool builders beyond the original vendor to target or simulate the hardware correctly, or for algorithm
designers to be confident in their code. To address these gaps, prior studies have probed the behavior of these units with manually crafted tests. Such tests are cumbersome to design, and adapting them as the accelerators evolve requires repeated manual effort.

We present a formal model for the tensor cores of
Nvidia's Volta, Turing, and Ampere GPUs.  We identify specific
properties---rounding mode, precision, and accumulation order---that
drive these cores' behavior.  We formalize these properties and then use
the formalization to automatically generate discriminating inputs that
illustrate differences among machines.  Our results confirm many of the
findings of previous tensor core studies, but also identify subtle
disagreements.  In particular, Nvidia's machines do not, as previously
reported, use round-to-zero for accumulation, and their 5-term
accumulator requires 3 extra carry-out bits for full accuracy.  Using
our formal model, we analyze two existing algorithms that use
half-precision tensor cores to accelerate single-precision
multiplication with error correction.  Our analysis reveals that the
newer algorithm, designed to be more accurate than the first, is
actually less accurate for certain inputs.
\end{abstract}

\section{Introduction}\label{sec:intro}
As applications strive for greater performance in today's post Moore's Law era, hardware designers have turned to specialized hardware units and non-standard ISA extensions to satisfy this need. In 2016, Google announced its Tensor Processing Unit, a hardware unit specializing in matrix multiplication \cite{google-tpu}. The next year, NVIDIA announced its new Volta architecture would feature Tensor Cores \cite{volta-whitepaper} which have since evolved with each new architecture.
Today, all major CPU and GPU vendors incorporate ISA extensions for matrix multiplication that operate on reduced-precision floating point formats.
Notably, the functionality of these extensions is not standardized yet, and is poorly documented as well.
While non-standard designs might be successfully employed within closed-source vendor libraries and for low-precision AI applications, they can prove to be difficult and error-prone for others who seek to build innovative linear algebra methods that are precision-sensitive~\cite{haidar2020mixed}.
Tensor cores have demonstrated numerical inconsistency across architectures that has an impact on the portability of algorithms. For example, a tensor core implementation of Fast-Fourier transform~\cite{tcfft2024} saw its mean relative error drop by as much as 34\\footnote{An accuracy test provided that accompanies the code reports an error of $1.5e^{-2}$ on Volta and $9.92e^{-3}$ on Ampere} when moving from Volta's tensor cores to Ampere's. This shifting behavior, coupled with the lack of a specification, presents a challenge for safety-critical applications that are sensitive to an implementation's numerical behavior. Furthermore, without a behavioral specification, efforts such as \citet{Goodloe_Muñoz_Kirchner_Correnson_2013} that verify numerical programs cannot be utilized.
Thankfully, work done by researchers to understand tensor core functionality~\cite{fasi2021numerical,Hickmann2019ExperimentalAO,xinyili24} has led to novel algorithms that can use these cores to speed up even single-precision computations for HPC~\cite{markidis,ootomo2022recovering}. As new cores are released, though, these same efforts must be repeated.

In this paper, we provide a formal description for the tensor cores across three generations of graphics cards. These formal descriptions can not only provide accurate and reliable component-level specifications enabling automated reasoning about their functionality but also facilitate the the creation of novel, hitherto unimagined, uses, while also improving debuggability, correctness checking, and security analyses.

Our models of tensor cores support two key properties: i)~they are executable, ii)~they can be used in automated reasoning. Our models are also parametric, enabling them to be quickly adapted for new architectures.

Our novel contributions include:

\begin{itemize}
\item A formalization of the numerical properties of mixed-precision block FMA, collected from prior literature, that can be used to identify the properties of different matrix multiplication units
\item A formal, executable model of the matrix multiplication units across three generations of GPUs - Volta, Turing, and Ampere
\item A revision of a mischaracterization in prior work that had concluded the rounding mode of tensor cores is round-to-zero. The actual rounding behavior, truncation, is subtly different. 
\item An analysis of two error-correcting matrix-multiplication algorithms that shows, due to the properties of the tensor cores, the method which trades speed for accuracy can actually produce less accurate results than its faster counterpart.
\end{itemize}

The functional and performance aspects of tensor core behavior have received significant scrutiny through testing~\cite{sun2022dissecting,fasi2021numerical,yan2020demystifying}.
Nevertheless, our SMT formalization unearths subtle discrepancies between test-based reverse engineered descriptions and the actual hardware.

In addition to enabling program analyses, formal models enable the construction of hardware simulators. These simulators in turn allow developers to evaluate the numerical behavior of their algorithms on multiple different architectures, all from the same machine and without requiring access to many different and potentially expensive devices. 

The rest of this paper is structured as follows. 
Section~\ref{sec:relwork}
provides a list of closely related work.
Section~\ref{sec:ptx-sass} provides
background on PTX and SASS, two instruction sets pertinent to
NVIDIA GPUs and necessary to understand Tensor Cores. It also details the HMMA instruction that carries out the matrix
operation $D=A\times B+C$.
Then, in section \ref{sec:fasi-case-study}, we formalize the numerical properties of tensor cores, compare our findings with previous works, and describe our resulting formal model. In section \ref{sec:ootomo-case-study}, we study two methods used to perform single-precision multiplications using the half-precision tensor cores discussed in \citet{ootomo2022recovering}.  We then analyze these algorithms, using SMT to try to prove that the error of Ootomo and Yokota's method is always better than \citet{markidis}.

\label{sec:bg}
\section{Related Work}
\label{sec:relwork}
We survey closely related work on floating-point formalization and testing-based specification discovery, followed by some non-floating-point formalization efforts.

An SMT theory for floating point reasoning was proposed by \citet{rummer2010smt}, which also included formalizations for rounding modes.  However, SMT-based floating point reasoning has historically been found to have poor scalability~\cite{darulova2014sound,schkufza2014stochastic}, but has been successfully used for error analysis~\cite{solovyev2018rigorous}. \citet{leeser2014make} demonstrated success in using SMT for floating-point reasoning, albeit using the theory of Reals.  \citet{martin2015automatable} redefined the floating point theory, substantially improving SMT's capabilities.  \citet{darulova2018daisy} used SMT to statically analyze floating point programs, for instance to compare roundoff errors between fixed-point and floating-point arithmetic. Floating point capabilities have similarly been implemented in other theorem provers, such as Coq~\cite{2011flcoq} which has also been used for error analysis~\cite{appel2024vcfloat2}. Each of these formalizations follows the IEEE standard~\cite{ieee-754-2008} and hence do not contain support for the non-standard accumulator which our work provides.
Titolo et al.~\cite{laura-titolo-munoz-vmcai} present an abstract interpretation framework for floating-point program roundoff error analysis.

Using tests to identify the implementation peculiarities of floating point units dates as far back as~\citet{paranoia}.  
In the case of GPU tensor cores, there has been considerable interest in understanding their functional as well as performance characteristics. \citet{sun2022dissecting} studied the tensor core implementations across various NVIDIA architectures.  While they primarily focused on the throughput and latency, they briefly investigated the numerical behavior of tensor cores by studying the relative error for different floating point formats. \citet{blanchard2020mixed} devised a framework to perform an error analysis of block fused multiply-add units.  Their method incorporates the supported precision of the uni in its formulation, allowing it to support future units that may offer a different precision. \citet{Hickmann2019ExperimentalAO} and \citet{fasi2021numerical} studied tensor cores by using carefully constructed experiments to determine the hardware's behavior such as its rounding mode, precision, and support for subnormals. Xinyi et al.~\cite{xinyili24} employed similar techniques while further exploring the block-FMA feature and additional bits for tensor cores and AMD's matrix cores.  \citet{yan2020demystifying} also studied the instruction-level details of the tensor cores, providing insights into how the matrix operation is performed on Turing, showing how the threads in a warp cooperate to compute the mma operation.

Formal descriptions of architectural components have been used to detect subtle correctness and security properties unrelated to floating-point arithmetic.
The Check tools (TriCheck~\cite{trippel2017tricheck}, CoatCheck~\cite{lustig_coatcheck_2016}, 
CCICheck~\cite{manerkar_ccicheck_2015}, 
PipeCheck~\cite{lustig_pipecheck_2014}), focus on memory consistency models and highlight the pitfalls resulting from under-specified ISA details.
The CheckMate tool~\cite{trippel_checkmate_2018} uses model checking to automatically create exploits for cache side channels.
Manual formalization of specifications is costly and this has led to work that seeks to automate the creation of formal ISA semantics. SAIL~\cite{armstrong_isa_2019} and K~\cite{dasgupta_complete_2019} have been explicitly built for ISA specifications. For x86, \citet{godefroid-taly} leveraged SMT to find input examples, while ~\citet{heule_stratified_2016} explored stratified synthesis. Using program synthesis has been explored to automatically formalize hardware specifications for memory consistency models~\cite{hsiao2021synthesizing,norman2023pipesynth}.

\section{PTX and SASS Background}
\label{sec:ptx-sass}

The NVIDIA GPUs we use in this work are commonly programmed in the CUDA programming language, a C++ dialect that supports explicit data parallelism and the ability to specify which functions run on the CPU and which run on the GPU.
To use the tensor cores, CUDA provides library functions that are internally implemented using inline assembly in the \textit{virtual} PTX instruction set architecture (ISA)~\cite{nvidia-ptx}.
PTX is a GPU independent ISA which resembles a compiler intermediate representation with features such as types, infinite registers, scoping, and so on. 
The physical ISA, commonly referred to as SASS~\cite{BinaryUtilities}, resembles a more traditional machine ISA and, unlike PTX, changes across GPU architectures.
PTX is compiled to SASS using an architecture-specific assembler called \texttt{ptxas}.
PTX provides forward compatibility with newer GPU architectures. 
If the GPU driver does not find the SASS for the current architecture in the executable, it will recompile the PTX in the executable at runtime to the architecture-appropriate SASS.
While NVIDIA provides a PTX specification, it does not provide any information about SASS, prompting reverse engineering efforts~\cite{sun2022dissecting, yan2020demystifying, fang2022towards,hayes_decoding_2019}.

\subsection{Tensor Cores and the \texttt{HMMA} Instruction}
\label{sec:tcbg}

\texttt{HMMA} is the primary SASS instruction that interacts with the tensor cores~\cite{jia2018dissecting}.
Programmers usually use CUDA's Warp Matrix or \texttt{wmma} Functions~\cite[\S7.24]{cuda-c-programming} to use the tensor cores.
However, the cores can also be accessed directly using inline assembly by using the PTX \texttt{mma.m8n8k4} instruction. 
Examining with \texttt{cuobjdump}~\cite{BinaryUtilities} the disassembly of SASS programs that use either of these methods confirms variants of the \texttt{HMMA} SASS instruction are used.

Across different architectures, the behavior of the tensor cores and its corresponding \texttt{HMMA} instruction changes. On Volta and Turing, the tensor cores are invoked via the \texttt{HMMA.884} instruction, while the Ampere architecture replaced this with \texttt{HMMA.16816}. Both instructions multiply two matrices, $A$ and $B$, and add a third matrix, $C$, though the \texttt{884} variant operates on $4\times4$ matrices, while the \texttt{16816} variant operates on $8\times8$ matrices. In fact, this change highlights another portability concern: the \texttt{mma.m8n8k4} PTX instruction that previously invoked tensor cores no longer does so on the Ampere architecture. Instead, it produces a sequence of Fused Multiply-Add (\texttt{FMA}) instructions that use the device's slower floating point cores whose numerical properties are quite different from tensor cores.

For both the \texttt{884} and \texttt{16816} variants, the \texttt{HMMA} operation consists of three steps: 1) multiplying matrix A and B, 2) accumulating the products of A, B along with matrix C, and 3) rounding the final result.  Each element in the resulting $N\times N$ matrix $D$ is computed via the following equation:

\begin{equation}\label{eq:matrix-mul-formula}
\begin{split}
    D_{i,j} = A_{i,1}\cdot B_{1,j} + A_{i+1,1}\cdot &B_{1,j+1} + \ldots + A_{N, 1}\cdot B_{1, N}
    \end{split}
\end{equation}

Unlike most GPU instructions, where each thread's calculations are independent of other threads, the \texttt{HMMA} instruction requires all threads within the warp to cooperate to compute the result, and only one matrix multiplication is performed per warp per instruction. Prior work by \citet{yan2020demystifying,fang2022towards} has described how matrix elements are mapped to each participating threads' registers.
The \texttt{HMMA} instruction supports both F16 and F32 types for elements of $C$ and $D$ which can be individually specified. $A$ and $B$ are always F16. Since Volta, tensor cores introduced have support for more formats: INT4 and INT8 in Turing, followed by Ampere's support for double-precision (FP64) and a custom format, Tensor Float 32 (TF32). This work focuses on the FP16 and FP32 formats that are supported on all tensor cores, though the properties we establish can be adapted to study tensor core behavior for different formats.

\section{Tensor Core Semantics}
\label{sec:fasi-case-study}
Although Equation~\ref{eq:matrix-mul-formula} appears to be a sufficient description of how \texttt{HMMA} behaves, floating point cognoscenti will immediately inquire about the following details which are needed to build a sufficiently detailed formal model:    

\begin{enumerate}    
    \item Are the multiplications and additions exact? What rounding mode is then used?
    \item Since standard floating point addition is not associative, how does the computation differ when terms are rearranged?
    \item Are the intermediate sums normalized, or only the final result?
                    \end{enumerate}

NVIDIA detailed the architecture of their tensor cores in their whitepaper describing the Volta GPUs~\cite{volta-whitepaper}, though is missing this level of detail:
\begin{quote}Tensor Cores operate on FP16 input data with FP32 accumulation. The FP16 multiply results in a full precision product that is then accumulated using FP32 addition with the other intermediate products for a $4\times4\times4$ matrix multiply.
\end{quote}

The PTX documentation is also unhelpful, stating that ``The accumulation order, rounding, and handling of subnormal inputs is unspecified.'' \cite[\S9.7.13.4.14]{ptx-isa-73}.  Previously, \citet{fasi2021numerical} answered some of these questions for the Volta architecture using 
carefully reasoned empirical tests
Our goal, in contrast, is to provide a complete description of tensor core behavior
as a formal model to not only answer such questions but also allow reasoning about other properties.

To establish a precise semantics for the \texttt{HMMA} instruction and its variants, we focus on only the first row and column of matrices \texttt{A} and \texttt{B} and the first element of
matrix \texttt{C}, making the instruction equivalent to equation \ref{eq:matrix-mul-formula} when $i=j=1$.

The operations in equation \ref{eq:matrix-mul-formula} may be implemented in myriad ways impacting the final result. Constructing a model requires determining which choices were made in hardware. Doing this unavoidably requires some manual effort to model the hardware design space and inform the possible implementations that need to be evaluated.  To construct our model, we scoured the literature for implementations of matrix multiplication and dot products.  The specification that accumulation is done in FP32 enables us to eliminate several possibilities, namely techniques like \citet{bohlender2012fast} for exact accumulation and the fixed-point approaches such as described by \citet{boldo:hal-02982017}. Ultimately, our tensor core model is primarily built upon the existing work of \citet{fasi2021numerical} and \citet{Hickmann2019ExperimentalAO}. We revise their findings and offer models for Turing an Ampere with a framework that can quickly adapt to new architectures. 

Like previous work, we use tests to discriminate between the different possibilities.
However, unlike previous work, these discriminating tests are {\em automatically generated}
using an automated theorem prover (e.g., cvc5~\cite{cvc5}).
Essentially, we write formulae to capture the possible ways in which the implementation of the tensor core unit may behave {\em and ask the automated theorem prover to find input values that would yield different outputs based on the design choices under investigation.} Our framework consists of these formalizations, encoded in SMT, and produces inputs that are then provided to the hardware to probe their behavior. The queries are fully parametric, allowing them to be easily adjusted to probe different implementation possibilities.

Using SMT solvers to generate inputs this way is routine~\cite{peleska2011automated,kim2019test}, but we are not aware of prior work that utilizes them to elicit the latent numeric behavior of nonstandard floating point operations that are performed by GPU tensor cores.  This approach is particularly well suited for the ill-documented tensor cores whose behavior continues to evolve with each new generation. New tests can easily be automatically generated as the underlying architecture changes. For instance, the tensor cores in Volta and Turing multiplied $4\times4$ matrices. Ampere shifted to $8\times8$ matrices, requiring new tests to explore its behavior. 

Additionally, compared to prior work that has investigated these properties without using SMT, we contribute knowledge about corner-cases involving rounding-modes and the number of carry bits required due to lack of normalization.\footnote{Recent growth
in power of SMT-solvers to deal with floating-point queries was essential for this to be practical~\cite{brain_building_2019}; see Table~\ref{tab:query-timings} and surrounding discussions.}

\subsection{Precision}
To demonstrate our approach, we begin by testing one of the claims made in the whitepaper: that the multiplication between the elements of $A$ and $B$ are performed in single-precision. We use SMT to identify values that are representable in FP16 but whose product is not.  Then, we provide these values to the hardware and determine whether the correct result is reported.

\begin{figure}[!t]
\begin{lstlisting}[style=mystyle, language=Python,label={lst:example-proof},caption={The python script showing how to use cvc5's (or Z3's) python api to identify a pair of half-precision values whose product is not exact when the multiplication is performed in half-precision.}]
from cvc5.pythonic import *  # can also import Z3
s = Solver()
a = FreshConst(Float16())
b = FreshConst(Float16())
for rm in {RNE(), RTZ(), RTN(), RTP()}:
    fp16_result = fpToFP(RTZ(), fpMul(rm, a, b), Float32()) (*\label{code:fp16-line}*)
    fp32_result = fpMul(rm, fpToFP(RTZ(), a, Float32()), fpToFP(RTZ(), b, Float32()) (*\label{code:fp32-line}*)
    s.add(Not(fp16_result == fp32_result)) (*\label{code:proof-line}*)
if s.check() == sat:(*\label{code:check-line}*)
    m = s.get_model()(*\label{code:model-line}*)
    # record values for a and b
    print(m.eval(a), m.eval(b)) (*\label{code:print-line}*)
else:
    print("Unsat/unknown")
\end{lstlisting}
\end{figure}

Listing \ref{lst:example-proof} demonstrates how we identify discriminating inputs for our tests using cvc5's Python API \cite{cvc5}.
Lines \ref{code:fp16-line} and \ref{code:fp32-line} express the multiplication in half-precision and single-precision respectively, with the for loop iterating over each rounding mode. Line \ref{code:proof-line} asserts that the two results differ.
Each of these lines adds a constraint on the values of $a$ and $b$ that must be met in order for the model to be satisfiable.  In line \ref{code:check-line}, we ask the solver to find values for $a$ and $b$ which satisfy each of these conditions.  A \texttt{sat} response indicates the solver has found such values.  In line \ref{code:model-line}, \texttt{get\_model} obtains these values from the solver, producing the values shown in Table \ref{tab:prop-1}.
 Once the solver has proved that the assertion holds, we can then extract the model in order to obtain inputs which can test if the tensor cores indeed perform the multiplication in single-precision (line \ref{code:print-line}).
\begin{table}
 \centering
 \small
 \caption{Values showing the multiplications are performed with full precision}
 \label{tab:prop-1}
 \begin{tabular}{|l|l|}
    \hline
      \textbf{a} & $1.2587890625\cdot2^{\text{-}15}$ \\
      \textbf{b} & $1.3681640625\cdot2^{\text{-}1}$\\
      \textbf{Exact Result} & $1.4162635803222656\cdot2^{\text{-}17}$\\
      \textbf{Result (half)} & $1.1767578125\cdot2^{\text{-}15}$ \\
      \textbf{Hardware Result} & $1.4162635803222656\cdot2^{\text{-}17}$\\
      \hline
 \end{tabular}
\end{table}

\citet{fasi2021numerical} presumably used
manual analysis in order to identify inputs that could be used to test the numerical behavior of the tensor cores such as its rounding mode and support for subnormals.  Here, we show that an automated theorem prover can be used to avoid this manual effort.
In addition, our method found two discrepancies in the work by \citeauthor{fasi2021numerical}, which we elaborate on in detail in the following sections. 

\begin{table}
\centering
\caption{\label{tab:notation}A description of the notation and terms used in the properties}
\begin{tabularx}{\linewidth}{lX}
\hline
\textbf{Symbol} & \textbf{Meaning} \\
$a~\cdot~b$ & Denotes multiplication of a and b in single-precision. \\
$a~\oplus_{rm}~b$ & Denotes addition of a and b in single-precision. \textit{rm}, when specified, denotes the rounding mode, defaulting to round-to-zero. \\
ToFP32(a) & Converts the half-precision input argument to its single-precision representation \\
ToFP16(rm, a) & Converts the single-precision input argument to half-precision, rounded with \textit{rm} \\
\hline
\end{tabularx}
\end{table}

\paragraph{Precision of Accumulation}
Tensor cores allow for the source and destination to hold values in either half precision or single precision.  The whitepaper states that the accumulation is performed in single-precision.  
The PTX documentation states that for half precision inputs and outputs, the accumulation is performed with ``at least half precision''~\cite[\S9.7.13.4.14]{ptx-isa-73}.
Here, we determine the actual precision used during accumulation for half-precision inputs and outputs by finding input values that satisfy the following:

\begin{myprop}\label{thm:accum-prec}
    $a \cdot b \oplus c \cdot d~\neq~\text{ToFP32}(a)~\cdot~\text{ToFP32}(b)~\oplus$ 
    $\text{ToFP32}(c)~\cdot~\text{ToFP32}(d)$\\
    Where $a, b, c, d, a\cdot b \oplus c\cdot d \in$ FP16, $a \cdot b, c \cdot d \not\in$ FP16
\end{myprop}

This states that certain FP16 inputs can yield products that cannot be represented in FP16, but whose sum can be. 
Our tests using inputs generated from the SMT solver 
confirm that the hardware performs accumulation in single-precision for half-precision inputs and outputs.

\subsection{Rounding}

Previously, \citet{fasi2021numerical} determined that the intermediate results of the accumulation were rounded using round-to-zero, while \citet{Hickmann2019ExperimentalAO} suggested that the results are truncated.
\citeauthor{fasi2021numerical} also determined that no additional bits were used for rounding. However, properly rounding RTZ requires additional guard bits.
In a standard floating point addition algorithm, the terms are aligned so that they have the same exponent.  This requires shifting the mantissa of the term with the smaller magnitude to the right in what is called the significand alignment step.  IEEE-754 \cite{ieee-754-2008} round-to-zero requires that the result be equal to the largest magnitude no larger than the exact result.  

\begin{table}
\caption{Round-to-zero demonstration}
\label{tab:rtz-rounding}
\begin{tabular}{|c|c|c|c|}
\hline
\textbf{Input a} & \textbf{Input c} & \textbf{Properly-rounded RTZ result} & \textbf{Tensor Core Result} \\
$2^1$ & $\text{-}2^{\text{-}40}$ & $2 - 2^{\text{-}23}$ & $2^1$ \hspace{1em} \xmark \\
\hline
\end{tabular}
\centering
\end{table}
Consider the example in table \ref{tab:rtz-rounding}. If $a$ and $c$ are $2^1$ and $\text{-}2^{\text{-}40}$ (and $b$ is one), then the result in round-to-zero mode should return the value $2-2^{\text{-}23}$.  However, because the significance alignment step requires shifting the $2^{\text{-}40}$ term 41 bits to the right, all of its bits would be lost and the result would be $2^{1}$. 

To properly handle this, floating point adders often make use of a ``sticky bit'' that tracks whether any bits were lost during alignment. However, a sticky bit does not work when aligning more than 2 terms, as the lost bits might have had different magnitudes across terms. Instead, accurate RTZ rounding requires preserving the bits that would be lost during the alignment, which \citeauthor{fasi2021numerical} concluded were not present in Volta.

Our tests for the rounding mode on hardware generate inputs that discriminate between each pair of rounding modes.\footnote{Available online at \url{https://pyxis-roc.github.io/tensor\_core\_semantics/}} Evaluating the resulting inputs, we find that tensor cores do not adhere to any of the IEEE-754 rounding modes, {\em including}  RTZ rounding, contrary to the findings in \citet{fasi2021numerical}. Instead, they truncate. On the Volta and Turing tensor cores, adding the aforementioned example, $2^1$ and $\text{-}2^{\text{-}40}$, results in $2^1$, which would be the result in round-to-nearest. While one may conclude that the rounding mode depends on the inputs, there is in fact a simpler explanation: during the significand alignment step, mantissa bits that were discarded during the shift are lost.  This means that performing an effective subtraction with numbers having an exponent difference greater than the number of bits in the mantissa (23 in this case) is the same as adding zero.  As mentioned before, this behavior violates the guarantees of round-to-zero which mandates the rounded result cannot be greater than the true result. This difference only manifests in effective subtraction, a subtlety that explains the mischaracterization while exposing the fragility of informal tests.

Tensor cores support outputting a FP16 result, requiring the FP32 accumulation to be rounded. We encode property \ref{thm:round-fp16} and generate tests that identify the rounding mode used in this case.
 
\begin{myprop}\label{thm:round-fp16}
    ToFP16 $\big(rm_1,~ToFP32(a)~\cdot~ToFP32(b)\big)~\neq $ \\
    $ToFP16 \big(rm_2,~\text{ToFP32}(a)~\cdot~\text{ToFP32}(b)\big)$ \hspace{2em}
    Where $a,b \in$ FP16, $rm_1 \neq rm_2$
\end{myprop}
In this experiment, we find a pair of values in FP16 whose product, when rounded to FP16, differs for different rounding modes.  We limit ourselves to a single term, setting the rest of the values to 0 so as to avoid behavior that may be attributed to the rounding mode that is used for computing the partial sums. From the experiments, we conclude that the final rounding is performed in round-to-nearest. While the behavior is consistent with the findings in \citeauthor{fasi2021numerical}, there it was concluded that this rounding was done in software. Our experiments reveal that this is not the case. This is further evidenced upon examining the disassembly (listing \ref{lst:hmma-disasm}) which shows no intervening instructions before the result is stored to memory.\footnote{This disassembly is consistent across multiple different compiler versions --- we tested nvcc versions 11.3 through 12.4} This indicates that the rounding to fp16 is in fact handled by the tensor cores.

\begin{lstlisting}[label={lst:hmma-disasm},caption={The SASS disassembly (Volta) for half-precision mma. Only the first store is shown.},basicstyle=\ttfamily\footnotesize,captionpos=b,float,frame=single,breaklines=true]
HMMA.884.F16.F16.STEP0 R12, R32.ROW, R2.COL, R12 ;
HMMA.884.F16.F16.STEP1 R14, R32.ROW, R2.COL, R14 ;
ST.E.SYS [R4], R12 ; /* ... */
\end{lstlisting}

\subsection{Accumulation Order \& Normalization}
IEEE-754 addition is not associative due to the rounding and normalization that occurs after each operation. \citet{fasi2021numerical} determined that the accumulation for Volta is performed into the element with the largest magnitude, and that there is no normalization of intermediate sums. We evaluate property \ref{thm:accumulation-order} and determine that the result does not depend on the order of the terms. 

\begin{myprop}\label{thm:accumulation-order}
    $(a_1 b_1~\oplus_{rtz}~a_2  b_2 )~\oplus_{rtz}~a_3\cdot b_3 \neq a_1 \cdot b_1~\oplus_{rtz}~(a_2 \cdot b_2 $
    $\oplus_{rtz}~a_3\cdot b_3)$
\end{myprop}

Testing for normalization requires an implementation of a multi-term floating-point accumulator that could accumulate without normalizing intermediate sums. However, the floating-point operations provided by SMTLIB are IEEE-754 compliant, which means that the intermediate sums will always be normalized. To overcome this, we developed our own floating-point accumulator using bitvector operations.  Our implementation takes into account each of the previous discoveries regarding tensor cores.

\begin{myprop}\label{thm:normalization}
$(a_1\cdot b_1~\oplus~a_2\cdot b_2)~\oplus~c~\neq$
no-normalize-sum($a_1\cdot b_1$, ~$a_2\cdot b_2,~c$).\\
Where $a_1,a_2,b_1,b_2, c > 0$ and no-normalize-sum(x, y, z) sums x, y, and z without normalizing the intermediate results.
\end{myprop}

To accumulate without normalization, the terms are first aligned to the maximal exponent before being accumulated. Terms are aligned by right shifting their mantissas according to the difference between their exponent and the maximal exponent. An implementation may choose whether or not to track some or all of the bits that were shifted out. We find that all shifted bits are discarded on Volta and Turing, while one is preserved on Ampere.  Evaluating the inputs provided by the theorem prover for property \ref{thm:normalization} confirms that the tensor cores do not normalize intermediate sums. 
\paragraph{Number of Carry-Out Bits Required Due to Lack of Normalization} As \citet{fasi2021numerical} noted, for accumulation of N-terms, $\lceil\log_2(N)\rceil$ extra carry-out bits are needed, meaning 5-terms require 3 bits of carry out. However, they were only able to find examples that required 2 carry-out bits. 
Finding inputs that show that 3 extra carry-out bits are needed (and are used by the actual tensor cores) in order to perform 5-term accumulation is a perfect task for SMT solvers.  As our implementation is fully parametric, we can easily model the use of differing numbers of carry bits and then use SMT to find values where they differ. It takes cvc5 just over a minute to find inputs proving that Volta and Turing require 3 bits for carry-out, which we confirm by finding the hardware computes the correct result.

When we evaluate the values produced by the automated theorem prover, we find that the hardware reports the value that would be the result if 3 bits were indeed used.  Additionally, we also use SMT to prove that no more than 3 bits are needed by proving that accumulation with 4 extra bits is equivalent to accumulation with 3 extra bits. We thus improve upon the findings of \citet{fasi2021numerical} and find that the 5-term accumulator uses 3 extra bits for carry out. Ampere's 9-term accumulator should likewise require 4 extra bits. However, attempts with both cvc5 and Z3 to find inputs to confirm this timed out after 6 hours.

\section{Model and Results}

\begin{table}
\caption{\label{tab:query-timings}Timings for the queries sent to the solvers. In the last row, - means that the solvers were unable to find a result given a 6 hour timeout}
\centering
\begin{tabular}{lrr}
    \hline
    \textbf{Query} & \textbf{Z3} & \textbf{cvc5} \\
    \hline
    Exact Multiplication & 0.55s & 0.027s \\
    Exact Addition in FP16 & 27.59s & 0.12s \\
    Rounding of Final Result & 0.52s & 0.008s\\
    Rounding of Accumulator & 0.34s & 0.017s \\ 
    Accumulation Order & 9.87s & 0.25s \\
    Normalization & 9.63s & 0.25s \\
    3 Carry Bits & 410.25s & 72.793s \\
    4 Carry Bits (Ampere) & - & - \\
\hline
\end{tabular}
\end{table}

\begin{algorithm}[t]
\footnotesize
\begin{algorithmic}
\State  \textbf{Inputs:} Two four(eight)-term fp16 vectors $a$ and $b$ and one fp16/fp32 scalar $c$
\State  \textbf{Output:} One fp16/fp32 scalar
\State \textbf{1}: Pairwise-multiply $a$ and $b$, computing their exact results in fp32, to get $ab$
\State \textbf{2}: Collect $ab$ together with the fp32 representation of $c$ and find the largest exponent
\State \textbf{2b}: If any number is NaN, or there are infinities of different signs, then return NaN
\State \textbf{2c}: (Add one bit of padding to the least significant bit of each term's mantissa)
\State \textbf{3}: Right shift each term's 24(25)-bit mantissa, discarding all excess bits
\State \textbf{4}: Add three(four) bits of padding to the most significant bit of each term's mantissa
\State \textbf{5}: Take the now 27(29)-bit mantissas and accumulate them in any order
\State \textbf{6}: Normalize the result to fp32, shifting the mantissa and adjusting the exponent
\State \textbf{7}: Discard the upper 3(4) bits (and the lowest one bit)
\State \textbf{8}: Return the result in the requested precision, using round-to-nearest for fp16
\end{algorithmic}
\caption{Dot-Product on Volta, Turing, and Ampere. (Parts in parenthesis correspond to Ampere)}
\label{alg:mma}
\end{algorithm}

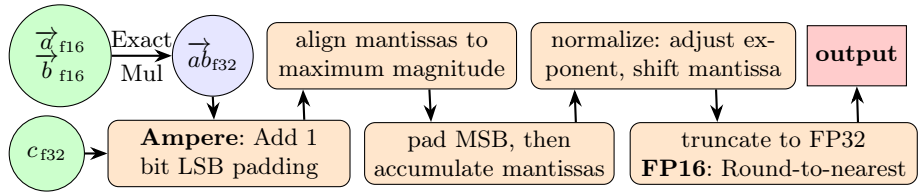
\begin{figure}[t]
\begin{tikzpicture}[
    intermediate/.style={circle, fill=blue!10, draw, text centered},
    process/.style={rectangle, draw, rounded corners, text centered, minimum width=2cm, minimum height=0.4cm, fill=orange!20},
    input/.style={circle, draw, text centered, minimum width=1cm, minimum height=0.8cm, fill=green!20},
    output/.style={rectangle, draw, fill=red!20, minimum height=0.8cm},
    arrow/.style={thick, ->, >=Stealth}
]

\node [text width = 0.8cm] (input) [input] {$\overrightarrow{a}_\text{f16}$ $\overrightarrow{b}_\text{f16}$};

\node (ab) [intermediate, right=0.8cm of input.east, anchor=west] {$\overrightarrow{ab}_\text{f32}$};
\node (c) [input, below=1.3cm of input.west, anchor = west] {$c_\text{\hspace{0.15em}f32}$};
\node [text width = 3cm] (pad) [process, right=0.3cm of c.east, anchor = west] {\textbf{Ampere}: Add 1 bit LSB padding};
\node [text width = 3.1cm] (align) [process, right=0.17cm of ab.east]  {align mantissas to \\ maximum magnitude};
\node [text width = 3.1cm] (accumulate) [process, right=0.18cm of pad.east, anchor=west] {pad MSB, then accumulate mantissas};
\node [text width = 3.3cm] (normalize) [process, right=0.18cm of align.east, anchor=west] {normalize: adjust exponent, shift mantissa};

\node [text width = 3.6cm] (round) [process, right=0.20cm of accumulate, anchor=west] {truncate to FP32 \\\textbf{FP16}: Round-to-nearest};
\node [output, right=0.14cm of normalize.east] (output) {\textbf{output}};

\draw [arrow] (input.east) -- node[midway, above] {Exact} node[midway, below] {Mul} (ab.west);
\draw [arrow] (c.east) -- (pad.west);
\draw [arrow] (ab.south) -- ([xshift=-0.2cm] pad.north);
\draw [arrow] ([xshift=1cm] pad.north) -- ([xshift=-1.18cm] align.south);
\draw [arrow] ([xshift=0.5cm] align.south) -- ([xshift=-0.775cm] accumulate.north);
\draw [arrow] ([xshift=1.15cm] accumulate.north) -- ([xshift=-1.18cm] normalize.south);
\draw [arrow] ([xshift=0.7cm] normalize.south) -- ([xshift = -0.75cm] round.north);
\draw [arrow] ([xshift=1.12cm] round.north) -- (output.south);
\end{tikzpicture}
\caption{Behavior of Tensor Cores}
\label{fig:tensor-core}
\end{figure}
Figure \ref{fig:tensor-core} outlines the behavior of the tensor cores' dot product step from equation \ref{eq:matrix-mul-formula}, which is described in more detail in algorithm \ref{alg:mma}.\footnote{The full implementation is available online at \url{https://github.com/pyxis-roc/tensor\_core\_semantics}} The inputs, (shown in green in the figure) are two fp16 vectors containing four (eight on Ampere) scalars, and an fp16 or fp32 scalar. Note that for the fp16 $a$ and $b$ values, the multiplication of their 11-bit mantissas can always be represented in fp32's 24 bit mantissa, meaning that no rounding is required and the results are always exact. Our findings show that the Volta and Turing tensor cores possess identical numerical behavior. Ampere tensor cores differ in that the multiplication is now performed for $8\times8$ matrices. Additionally, while Volta and Turing discard all shifted bits during significand alignment, on Ampere, one shifted bit is preserved.

Table \ref{tab:query-timings} demonstrates how the recent advancements in SMT's floating point capabilities have dramatically improved SMT's floating-point capabilities, resulting in impressive query times that in many cases respond in sub-second time. As expected, the more inputs that are required to prove the property, the longer it takes for the query. Queries for exact multiplication, addition and rounding each require just two inputs
\footnote{or four if querying for fp16 values. When possible, we asked the solver to first find fp32 values that demonstrated the property, and then asked for fp16 values whose product results in that value},
normalization and accumulation order require 3, and 3 carry bits requires 9 inputs. Proving Ampere's 9-term accumulator requires 4 carry bits requires finding 17 total inputs, a task which the current solvers were ultimately unable to finish.
Each of the queries are tested on a machine running an AMD EPYC 7502P 32-core processor with 256GB RAM at 1.5GHz and given a 6 hour timeout. For hardware tests, we used a Titan V GPU for Volta, an RTX 2080Ti for Turing, and an RTX A6000 for Ampere. Kernels were compiled with CUDA version 11.8. The timing results in Table \ref{tab:query-timings} show the query time, in seconds, using Z3 version 4.8.9~\cite{de_moura_z3_2008} and cvc5 version 1.0.2~\cite{cvc5}.

\section{Ootomo and Yokota Case Study}\label{sec:ootomo-case-study}

While tensor cores provide high-performance, their FP16 inputs mean they have low precision. 
Ingenious methods have therefore been developed to take advantage of tensor cores' performance without sacrificing precision.
One such method was proposed by \citet{markidis}, which introduced a residual matrix to record the loss of mantissa (the difference between FP32 and FP16 inputs) which can then be used to recover precision. A similar technique was used by \citet{fasi2023matrix} (which we abbreviate to O-Y). For a matrix product $A\cdot B$, the residual matrices $R_A$ and $R_B$ are calculated as the difference between the single-precision and half-precision representations of A and B as $R_A = A_{f32} - A_{f16}$ and $R_B = B_{f32} - B_{f16}$ respectively. The final recovered result is calculated using 
\begin{equation*}
A_{f32}\cdot B_{f32} = R_AR_B + A_{f16}R_B + R_AB_{f16} + A_{f16}B_{f16}
\end{equation*}

To further reduce the error, \citet{ootomo2022recovering} (which we abbreviate to O-Y) improved \citeauthor{markidis} method by incorporating rounding to nearest in the accumulator by performing accumulation outside of the tensor core unit. Additionally, they implemented scaling when computing the residual matrix. The updated procedure is as follows:

\begin{align*}
R_A &= (A_{f32} - A_{f16}) \cdot 2^{11} \text{\hspace{2em}}
R_B = (B_{f32} - B_{f16}) \cdot 2^{11} \\
A_{f32} B_{f32} &= \frac{R_AR_B}{2^{22}} + \frac{A_{f16}R_B + R_AB_{f16}}{2^{11}} + A_{f16}B_{f16} \\
D &= RN(A_{f32}B_{f32} + C)
\end{align*}
where $2^{11}$ is the scaling factor and RN denotes rounding to nearest.

The O-Y method is meant to improve the error correction of \citeauthor{markidis}'s at the cost of extra computation.  Using our models from Section~\ref{sec:fasi-case-study}, we implement both error correction methods described by the paper in SMT.  We then attempt to prove that the absolute accuracy for one of the final elements of the matrix, when using \cite{ootomo2022recovering} method, can never be worse than \cite{markidis}. To do this, we ask an SMT solver to prove the following formula:
\begin{myprop}
$\exists~\text{inputs}~s.t.~|Markidis(inputs) - actual(inputs)| < |Ootomo(inputs) - actual(inputs)|$ 
\end{myprop}
Where $Markidis$ and $Ootomo$ correspond to the result of one element in the final matrix computed using equation (6) and equation (24) from \citet{ootomo2022recovering}, respectively; $actual$ corresponds to the result obtained by performing the dot product in double precision. We also restrict the exponent ranges for the inputs to $2^{\text{-}15}$ and $2^{14}$ as was done for O-Y's Type 1 experiments.

Table \ref{tab:markidi-v-ootomo} shows the values for which O-Y's method has a higher error than \citeauthor{markidis}'s.  For this single query, it takes cvc5 less than 5 minutes
to find values for which the error using O-Y's method can be worse.  This is not to say that O-Y's method is worse overall, but rather proves that it is not more accurate for every input.  Nor does this contradict their empirical results showing that their method was more accurate in general.  In \citet{ootomo2022recovering}, it was noted that one of the main contributors to the error was due to the round-to-zero mode of tensor cores.  This means that for many cases, performing the accumulation outside of tensor cores can \emph{improve} the accuracy of the final result.  However, as we showed in section \ref{sec:fasi-case-study}, tensor cores do not normalize the intermediate sums.  This lack of normalization between each addition can improve the final accuracy by keeping parts of the mantissa that would have been lost during the normalization step thus making O-Y's method worse on some inputs.

\begin{table}
\centering
\small
\caption{\label{tab:markidi-v-ootomo}Inputs which show the the error of \citet{ootomo2022recovering} \textbf{(O-Y)} can be greater than \citet{markidis} (\textbf{M)}}
\begin{tabular}{|l|llll|}
\hline
$\vec{\textbf{a}}$ & $1.0009765625\cdot2^{\text{-}8}$ & $1.326171875\cdot2^{\text{-}14}$ & $2^{\text{-}12}$ & $2^{\text{-}12}$ \\
$\vec{\textbf{b}}$ & $1.998046875\cdot2^{\text{-}7}$ & $1.4443359375\cdot 2^{\text{-}7}$ &$2^{\text{-}12}$ & $2^{\text{-}12}$ \\
\textbf{\textit{c}} & \multicolumn{4}{c|}{$2^{\text{-}24}$} \\
\textbf{True} & \multicolumn{4}{c|}{$1 + 2^{\text{-}23}$} \\
\textbf{M} & \multicolumn{4}{c|}{$1 + 2^{\text{-}23}$} \\
\textbf{O-Y} & \multicolumn{4}{c|}{$1.0$} \\
\hline
\end{tabular}
\end{table}

The values produced by our experiment in Table \ref{tab:markidi-v-ootomo} follows the same pattern. To illustrate precisely why the error occurs, we walk through the example below.

\begin{enumerate}
    \item $\vec{a}$ and $\vec{b}$ are multiplied; largest exponent of all terms is -1.
    \item Each term is shifted to align their exponent to -1
    \item $a_1b_1$ and $a_2b_2$ are accumulated, resulting in exactly $1-2^{\text{-24}}$. \textit{This term is not normalized}.
    \item $2^{\text{-}24}$ ($a_3b_3$) is added to the previous term, resulting in exactly 1.0, represented internally as $2^{\text{-}1}\cdot 2^1$
    \item $2^{\text{-}24}$ ($a_4b_4$) is added to the previous term resulting in $1.0 + 2^{\text{-}24}$, represented as $2^{\text{-}1} \cdot 2^1 + 2^{\text{-}23}$
    \item At this point, O-Y's method diverges from \citeauthor{markidis}
    \begin{enumerate}
        \item 0.0 is added to $1.0 + 2^{\text{-}24}$
        \item $1.0 + 2^{\text{-}24}$ is normalized, yielding $1.0$, as the lowest bit is lost in the shift.
        \item $1.0$ is accumulated with $2^{\text{-}24}$ outside tensor cores. The result is $1.0$.
    \end{enumerate}
    \item In \citeauthor{markidis}, $c$ is accumulated \textit{inside} the tensor cores
    \begin{enumerate}
        \item $2^{\text{-}24}$ ($c$) is added to $1.0 + 2^{\text{-}24}$, resulting in $1.0 + 2^{\text{-}23}$, represented internally as $2^{\text{-}1}$ for the exponent with $2^1 + 2^{\text{-}22}$ in the mantissa.
        \item The term is normalized, yielding $1.0+2^{\text{-}23}$
    \end{enumerate}
\end{enumerate}
This experiment demonstrates precisely how the lack of normalization inside tensor cores can lead to a result with less error.
\citet[\S D-2]{fasi2021numerical} also demonstrated how normalization contributes to error with an experiment in which the value $1-2^{\text{-}24}$ is accumulated with four values, each $2^{\text{-}24}$.  When partial sums are normalized, the accumulation between $1-2^{\text{-}24}$ and $2^{\text{-}24}$ would result in the value of $1$. After being normalized, the exponent difference between the accumulated term and the remaining terms would cause the remaining additions to have no effect in round-to-zero, as their sums would be shifted out.  Instead, when the intermediate sum is not normalized, none of the bits from the $2^{\text{-}24}$ terms are lost and the final error is only $2^{\text{-}24}$.

\section{Conclusions and Future Work}
Using SMT, we formalized the properties of tensor cores and modeled their behavior across three generations.
We showed how the in-progress specification and an automated theorem prover could be used together to resolve contradictory observations obtained using solely test-based methods.
While most of our findings align with those of \citet{fasi2021numerical}, our model provided evidence that the rounding mode used for accumulation was simply truncation and that there 3 extra bits used for carry out for Volta and Turing's 5-term accumulator.
Once the model was built, we were able to use it and an automated theorem prover to investigate two algorithms that utilize tensor cores and examine claims about their relative accuracies, thus demonstrating the usefulness of our model to algorithm designers.

The framework we established is fully parametric and future work can reuse it to study the properties of tensor cores as they evolve with new generations. Preliminary experiments on Hopper GPUs (for which we lacked sufficient access to thoroughly study), for instance, indicate that even more bits may be preserved during significand alignment. Our model can also be adjusted to study different floating-point formats such as NVIDIA's 8-bit exponent, 10-bit mantissa TF32 format, or the two FP8 formats supported on Hopper. 

Given that future HPC hardware will likely be supported by non-standard hardware developed primarily for AI (including especially Tensor Cores)~\cite{reed2023hpc}, formalizations such as ours can play a central role in supporting reliable scientific computing in the future.
We plan to develop formal support to analyze such algorithms using techniques presented in this paper.

\section*{Acknowledgments}
This work is supported in part by NSF Awards 2403379, 2346394, 2217154, and 2124100.

\bibliographystyle{plainnat}
\bibliography{valpey-bibliography,ganesh,sree}

\clearpage
\onecolumn

\end{document}